\documentclass[aps,twocolumn,prl,tightenlines,floatfix,showpacs,10pt]{revtex4}
\usepackage[dvips]{graphicx}
\usepackage[english]{babel}
\usepackage{amssymb,amsmath} 
\usepackage{slashed}
\usepackage{hyperref}\usepackage{flushend}
\usepackage[caption=false]{subfig}

\hypersetup
  {citecolor = {blue},
  colorlinks = {true}, 
  urlcolor = {red}}

\def\mb{\mathbf}
\def\om{\omega}
\def\p{\partial}
\begin{document}
\title{Shear viscosity and imperfect fluidity in bosonic and fermionic superfluids}
\author{Rufus Boyack$^1$, Hao Guo$^2$ and K. Levin$^1$}
\affiliation{$^1$James Franck Institute, University of Chicago, Chicago, Illinois 60637, USA}
\affiliation{$^2$Department of Physics, Southeast University, Nanjing 211189, China}

\begin{abstract}
In this paper we address
the ratio of the shear viscosity to
entropy density $\eta/s$
in bosonic and fermionic superfluids. 
A small $\eta/s$ is associated with nearly perfect fluidity,
and more general measures of the fluidity perfection/imperfection
are of wide interest to a number of communities.
We use a Kubo approach to
concretely address this ratio via low temperature transport associated with
the quasi-particles.
Our analysis for bosonic superfluids  
utilizes the framework of the one-loop Bogoliubov approximation, 
whereas for fermionic superfluids we apply BCS theory and its BCS-BEC
extension.
Interestingly, we find that the transport properties 
of strict BCS and Bogoliubov superfluids
have very similar structures, 
albeit with different quasi-particle
dispersion relations.
While there is a dramatic contrast between the power law and exponential temperature
dependence for $\eta$ alone, the ratio
$\eta/s$ for both systems is more similar. 
Specifically we find the same linear dependence (on the ratio of temperature $T$ to inverse lifetime $\gamma(T)$) with
$\eta/s \propto T/\gamma(T)$, corresponding to imperfect fluidity.  
By contrast, near the unitary limit of BCS-BEC superfluids a very
different behavior results, which is more consistent with near-perfect fluidity.
\end{abstract}
\maketitle 
\vskip5mm

\section{Introduction} 
There is a growing interest among particle, condensed matter and atomic
physicists on the behavior of the shear viscosity $\eta$ and the degree to
which the ratio $\eta/s$, where $s$ is the entropy density, is close to
the lower bound $\eta/s\geq\hbar/(4\pi k_{B})$ 
conjectured by Kovtun, Son, and Starinets (KSS) \cite{Kovtun}. 
The KSS conjecture has led to renewed interest as to
which fluids in nature are ``perfect'' fluids, i.e., those that come as
close as possible to minimizing the conjectured bound.

For atomic Fermi gases there have been experimental \cite{ThomasI,ThomasII}
and theoretical \cite{GuoII} studies of this ratio which suggest a 
close approach to the KSS bound. These have been in
the specific unitary regime; no such studies are yet available
for the bosonic superfluids.
Theories of $\eta/s$ in graphene \cite{Fritz}, along with high $T_{C}$
superconductor experiments \cite{Rameau},
have also made claims that these exhibit an
$\eta/s$ near the KSS bound.

It is of interest then to perform concrete calculations of this ratio,
particularly in the presence of the many body physics which gives
rise to superfluidity. Thus, in this paper we address
the quasi-particle
contribution to $\eta/s$  
for bosonic and  fermionic superfluids. We arrive at a fairly generic
behavior for this ratio of the form
$\eta/s \propto T /\gamma(T)$, where $\gamma(T)$ is the 
temperature
dependent inverse quasi-particle lifetime.
While classical arguments have led to a prediction
of this form, we show that it holds as well for the extreme non-classical
regimes in systems with widely different quasi-particle dispersion relations.
A scaling of $\eta/s$ of this form has also been argued to apply near
quantum critical points, or for nodal $d$-wave superfluids, and in these
contexts, it has been argued that the behavior is
suggestive of near perfect fluidity \cite{Rameau}.

More generally,
here we present a comparison of the dissipative transport properties of bosonic
and fermionic superfluids mostly at low temperatures, where bosonic superfluid theories exist
and are controlled. For the fermionic superfluids there is no restriction on temperature, and one can, furthermore, 
probe the behavior of an interesting normal phase (with a pseudogap) in the regime of
BCS-BEC. In addition to the shear viscosity $\eta$, we focus
on the quasi-particle contributions to transport (which
are the exclusive contributions)
in the thermal conductivity $\kappa$, the $\omega \neq 0$
mass conductivity $\sigma$, and the off-diagonal
thermo-electric coefficients.
These are to be distinguished from condensate contributions,
which dominate the $\omega \equiv 0$ mass conductivity.
For the latter, a proper theory of transport has to deal with a number of
subtle features involving gauge invariance and the important
constraint in bosonic transport in which the (two particle) density
excitation spectrum or sound modes are intimately coupled to the single
particle excitations.
Our starting point for bosonic superfluids
is due to Wong and Gould \cite{Wong},
and Talbot and Griffin \cite{Griffin}, and known as the
``one-loop approximation''.

We note that prior to the recent focus on trapped atomic gas superfluids there
were puzzles concerning the behavior of the shear viscosity
alone, which originated in contrasting observations of the fermionic and
bosonic counterparts of liquid helium. 
For example, 
$^{3}\mathrm{He}$ and $^4\mathrm{He}$ exhibit a
remarkable difference in their shear viscosity for low
temperatures. At temperatures below the critical temperature
the shear viscosity of fermionic $^{3}\mathrm{He}$ has been 
measured \cite{Roobol} to be a decreasing function 
of decreasing temperature, whereas the shear
viscosity of bosonic $^4\mathrm{He}$ \cite{Woods} is an increasing function of
decreasing temperature. 
In this paper we suggest that these differences are understood
as reflecting the different dispersions of the quasiparticles. Indeed,
a central theme of this work is that the shear viscosity itself provides
a sensitive measure of
the nature of the quasi-particle excitations.

Our calculations show that the one-loop Bogoliubov theory 
for bosons and the BCS theory for fermions are
formally strikingly similar.
Nevertheless, primarily as a result
of differences in the quasi-particle excitation spectrum, as
well as the statistics, there
are important differences in superfluid
transport. Quite generally, in bosonic systems, because the dispersion relation is gapless, 
the transport coefficients increase more
rapidly as a function of temperature when compared to the gapped fermionic systems.
This results in low temperature transport in the bosonic case being more
accessible experimentally.

\section{Theory}
Previous studies of superfluid transport have
relied heavily on kinetic theory and a Boltzmann equation
ansatz \cite{Khalatnikov, GriffinBookI}. A less widely applied approach has
been the use of linear response theory and Kubo formulae,
which we will use here. 
The advantage of the Kubo formulae approach is that
by relating directly to Green's function diagrammatics one has better control over the
processes included in transport and the appropriate
constraints. This enables a more systematic imposition of perturbation expansions,
which is especially crucial when considering bosonic superfluids. Also important are the constraints which must
be imposed on separating the contributions associated with longitudinal and transverse correlations,
since it is only in the former that the condensate will directly enter.
Finally, a subtle but important issue here arises in the shear viscosity, for example, where the
Kubo formula shows that there are multiple response functions which enter in addition
to the simplest stress tensor-stress tensor correlator \cite{Bradlyn}. It is not as apparent how to include
these in a Boltzmann based approach.

Since it is likely
that dissipation in the ultracold gases
is linked to the details of the experimental set up, we will introduce dissipation
 via a phenomenological parametrization within the Kubo approach.
The philosophy behind our phenomenological approach 
to dissipation is similar to that articulated by Kadanoff and Martin,
who emphasized the importance 
of the Kubo based correlation functions
and their symmetries \cite{KadanoffMartinII}. 
In related work on superconductors \cite{KadanoffMartinI}, 
they argued for the suitability of introducing a 
parametrization of the lifetimes associated with transport.
In building any phenomenology it is important to emphasize that
inter-particle collisions can not be the sole source of dissipation
in mass transport, as in the particle conductivity. This
particular transport coefficient
reflects the fact that the total momentum would 
be conserved (in the presence of Galilean invariance)
without other sources of momentum relaxation.

By contrast, the strength of Boltzmann theory is that if all 
the details of the processes giving rise to dissipation are well
established, one can incorporate dissipation via
specific collision integrals. 
Within a Boltzmann based theory of bosonic superfluid transport,
there is a fairly extensive review 
by Griffin \cite{GriffinBookI} in which the shear viscosity and the thermal
conductivity are addressed. At a qualitative
level our Kubo calculations are consistent with this earlier work,
but we also include additional transport coefficients. A Kubo formulation
of the shear viscosity of a trapped Bose condensed gas 
was studied in \cite{Shahzamanian}  
within the second order Beliaev approximation, but
this analysis 
did not incorporate the contribution from anomalous Green's 
functions.
This paper revisits this earlier work (albeit without a trap)
and with the important inclusion
of the anomalous Green's functions, which are a crucial component for
a consistent treatment of superfluid transport.

\subsection{Physical Analysis of the Quasi-particle Regime}
In order to further understand the role of the phenomenological
inverse lifetime, we address $\eta/s$ using 
a simple classical argument \cite{KovtunII}.
The entropy density $s$ of a weakly interacting system is proportional
to the quasi-particle number density $n$:
\begin{align}s\sim k_B n.\nonumber\end{align}
The shear viscosity is proportional to the product of the average energy 
per particle $\epsilon$, and the mean free time between collisions $\tau \equiv 1/\gamma$:
\begin{align}\eta \sim n \epsilon/\gamma.\nonumber\end{align}
Then, assuming $\epsilon\sim k_B T$, the ratio $\eta/s$ is
\begin{align}\label{eq:EtaS}\eta/ s\sim T/\gamma.\end{align}
In order for the quasi-particle picture to be valid, the particles
must be long lived: $\hbar \gamma <<k_BT$ so that the ratio $\eta/s$ is far above the KSS bound:
$$\eta/s >>\hbar/4\pi k_B.$$
Importantly, we
will show in this paper that, even in the non-classical regime, for both 
bosonic and fermionic superfluids, an equation of the form given in Eq. (\ref{eq:EtaS}) results.
In this way the quasi-particle regime should be understood as a regime
where the system is far from being a perfect fluid.

\subsection{Overview of Our Transport Results} 
We begin by summarizing our results, which serve to emphasize
the formal similarity of the bosonic one-loop transport theory
with the fermionic BCS transport theory.
We define the general transport coefficients $(L_{ij})$ via particle $(\mb{J}_{p})$ 
and heat $(\mb{J}_{Q})$ current densities as follows
\begin{align}
\label{eq:Jp}\mb{J}_{p} &= -L_{11} \nabla \mu- L_{12} \nabla T, \\
\label{eq:JQ}\mb{J}_{Q} &= -L_{21} \nabla \mu- L_{22} \nabla T,
\end{align}
where $\nabla\mu$ and $\nabla T$ represent imposed gradients
of the chemical potential (analogous to the electric field for a
charged system) and the temperature. (We work in units where $\hbar=k_{B}=e=1$.)
Here the particle or mass conductivity $\sigma\equiv L_{11}$ and the
thermal conductivity $\kappa\equiv L_{22}$. The off-diagonal transport coefficients
appear, for example, in the quasi-particle thermopower.

For a superfluid it should be stressed that the various 
correlation functions that enter into the $L_{ij}$
may be distinct for longitudinal and transverse properties.
This distinction is most important for the mass conductivity,
as the longitudinal contribution reflects the condensate
(and diverges at zero momentum and frequency) while the
transverse contribution reflects the quasi-particles.
The shear viscosity is also represented in terms of this
transverse response.

Following the approach of Kadanoff and Martin \cite{KadanoffMartinI},
lifetime effects are phenomenologically incorporated by introducing
the parameter $\gamma(T)$. In this context $\gamma^{-1}$ was
introduced as a lifetime required to restore local equilibrium to a system 
perturbed from the equilibrium state. It may therefore be regarded as an additional experimental parameter
for the particular system of interest. In the context of superfluids $\gamma^{-1}$ 
can be associated with quasi-particle lifetime processes, which in certain cases
are known \cite{Khalatnikov}.

Using the correlation functions which will appear in Eqs. (\ref{eq:X11}$-$\ref{eq:X22})
below, we find that, for bosons, the 
transport coefficients (in 3d) are
\begin{align}\label{eq:NB}\eta^{B}&=\int_{0}^{\infty}dk\ 
\frac{k^6}{30\pi^2 m^2}\left(\frac{\xi_{\mb{k}}}{E_{\mb{k}}}\right)^2
\left(-\frac{\p n(E_{\mb{k}})}{\p E_{\mb{k}}}\right)\frac{1}{\gamma}, 
\\\label{eq:LB}\mathrm{Re}L^{B}_{ij}&=T^{1-j}\int_{0}^{\infty}dk\ 
\frac{k^{4}}{6\pi^2m^2}\xi_{\mb{k}}^{i+j-2}
\left(-\frac{\p n(E_{\mb{k}})}{\p E_{\mb{k}}}\right)\frac{1}{\gamma}.\end{align}
Note we have evaluated $\eta^{B}, \mathrm{Re}L^{B}_{ij}$ in the limit $\om\rightarrow0$.

We introduce $n_{0}$ as the condensate density and $g$ 
as the interaction strength. The Hugenholtz-Pines theorem determines
the chemical potential, in the Bogoliubov
approximation, as $\mu^{B}=n_{0}g$. 
The free particle dispersion relation is $\epsilon_{\mb{k}}=\tfrac{k^2}{2m}$
and we define $\xi_{\mb{k}}=\epsilon_{\mb{k}}+\mu^{B}$.
The Bogoliubov quasi-particle dispersion relation is then
$E_{\mb{k}}^2=\xi_{\mb{k}}^2-(\mu^{B})^2$.
We define $n(x)=[e^{x/T}-1]^{-1}$ as the Bose-Einstein distribution function.

The same calculations performed above for bosons can be performed 
for strict BCS fermions. The only differences that arise are 
a sign factor due to the different statistics, a degeneracy factor of two 
due to spin and a redefinition of the dispersion relation. 
For fermions the transport coefficients are
\begin{align}\label{eq:NF}\eta^{F}&=\int_{0}^{\infty}dk\ \frac{k^6}{15\pi^2 m^2}
\left(\frac{\xi_{\mb{k}}}{E_{\mb{k}}}\right)^2\left(-\frac{\p f(E_{\mb{k}})}{\p E_{\mb{k}}}\right)\frac{1}{\gamma}, 
\\\label{eq:LF}\mathrm{Re}L^{F}_{ij}&=T^{1-j}\int_{0}^{\infty}dk\ 
\frac{k^4}{3\pi^2m^2}\xi_{\mb{k}}^{i+j-2}\left(-\frac{\p f(E_{\mb{k}})}{\p E_{\mb{k}}}\right)\frac{1}{\gamma}.\end{align}
where $\xi_{\mb{k}}=\epsilon_{\mb{k}}-\mu^{F}$, $E_{\mb{k}}^2=\xi_{\mb{k}}^2+\Delta^2,$ and $f(x)=[e^{x/T}+1]^{-1}$.
(We have again evaluated $\eta^{F}, \mathrm{Re}L^{F}_{ij}$ in the limit $\om\rightarrow0$.) 
This expression for the shear viscosity has been obtained previously in \cite{GuoII}. 
Similarly the mass and thermal conductivities $L_{11}, L_{22}$ 
are consistent with results obtained from BCS theory \cite{KadanoffMartinI}. 
Our emphasis here is that a comparison between Eqs. (\ref{eq:NB}$-$\ref{eq:LB}) 
and Eqs. (\ref{eq:NF}$-$\ref{eq:LF}) shows the striking similarities 
between the transport coefficients in bosonic and fermionic superfluids.
A key difference between the transport coefficients
arises from the soft quasi-particle excitations for 
bosons as opposed to the gapped excitations for fermions.

\subsection{Details of the Derivation}  
We proceed now to derive 
Eqs. (\ref{eq:NB}$-$\ref{eq:LF}). 
In linear response theory
the response of a system perturbed slightly from
thermal equilibrium is expressed in terms of correlation
functions of the unperturbed system \cite{KadanoffMartinII}.
Equations (\ref{eq:Jp}$-$\ref{eq:JQ}) lead to four possible
correlation functions involving combinations of particle and heat or energy
currents.
These four
correlation functions are
\begin{align}\label{eq:X}\tensor{\chi}_{ij}(x_{1}-x_{2},\tau_{1}-\tau_{2})=-\langle
T_{\tau}j_{i}(x_{1},\tau_{1})j_{j}(x_{2},\tau_{2})\rangle,\end{align} where
$i,j\in\{1,2\}$.
The particle and heat currents which appear above
are defined as \cite{KadanoffMartinI} 
\begin{align}\label{eq:EC}j_{1}&=-\frac{i}{2m}\left(\nabla_{1}-\nabla_{1}'\right)\left.
\psi^{+}(1')\psi(1)\right|_{1'=1^{+}},
\\\label{eq:HC}j_{2}&=-\frac{i}{2m}\left(\p_{t_{2}}\nabla_{2}'+\p_{t_{2}}'\nabla_{2}\right)\left.
\psi^{+}(2')\psi(2)\right|_{2'=2^{+}}.\end{align}
For a generic superfluid correlation function $\tensor{\chi}_{ij}$, it is convenient to decompose into
longitudinal and transverse
components which are given by
$\chi_{ij}^{L}=\tfrac{\mb{q}\cdot
\tensor{\chi}_{ij}\cdot\mb{q}}{q^2},
\chi_{ij}^{T}=\tfrac{1}{2}(\sum_{\alpha}{\chi}^{\alpha\alpha}_{ij}
-\chi_{ij}^{L}).$

We define the Fourier transform by
$\tensor{\chi}_{ij}(x_{1}-x_{2},\tau_{1}-\tau_{2})
=\tfrac{1}{\beta}\sum_{i\om_{m}}\int\tfrac{d^{3}q}{(2\pi)^3}\tensor{\chi}_{ij}
(\mb{q},i\omega_{m})e^{i\mb{q}\cdot(\mb{x}_{1}-\mb{x}_{2})}e^{-i\om_{m}(\tau_{1}-
\tau_{2})}$.
Then the Kubo formulas for the transport coefficients, except those associated with $\chi_{11}$, are
\begin{align}\mathrm{Re}L_{ij}=-T^{1-j}\mathrm{lim}_{\mb{q}\rightarrow0}\frac{\mathrm{Im}\chi^{L}_
{ij}(\mb{q},\om)}{\omega},\quad i,j\neq1.\end{align}
Using this definition,  one can compute 
transport coefficients $\mathrm{Re}L^{B}_{ij}, i,j\neq1$
for the bosonic case and $\mathrm{Re}L^{F}_{ij}, i,j\neq1$ 
for the fermionic case.

The quasi-particle contribution to the mass conductivity and the shear
viscosity (for which there is no condensate
component) depend only on the transverse component of
$\tensor{\chi}_{11}$ and  are given by
\cite{KadanoffMartinII}
\begin{align}\mathrm{Re}\sigma(\om\neq0)&=-\mathrm{lim}_{\mb{q}\rightarrow0}
\frac{\mathrm{Im}\chi_{11}^{T}(\mb{q},\om)}{\om},
\\\label{eq:N}\eta&=-m^2\mathrm{lim}_{\om\rightarrow0}\mathrm{lim}_{\mb{q}\rightarrow0}
\frac{\om}{q^2}\mathrm{Im}\chi_{11}^{T}(\mb{q},\om).\end{align}
By limiting consideration in $\sigma$ to $\omega \neq 0$, we focus on the
quasi-particle transport. The total mass conductivity (which includes the condensate) is
$\mathrm{Re}\sigma(\om)=\mathrm{Re}\sigma(\om\neq0)+\frac{\pi
n_{s}}{m}\delta(\om),$ where $\tfrac{n_{s}}{m}$ is the superfluid
density. The mass conductivity of the condensate 
is infinite but all condensate thermoelectric coefficients vanish.
More specifically the condensate enters directly into only
$L_{11}$. Finally, we note that the
Onsager relation between the associated transport coefficients is
$L_{12}=L_{21}/T$.

\begin{figure*}
\subfloat{\includegraphics[width=2.25in,clip]{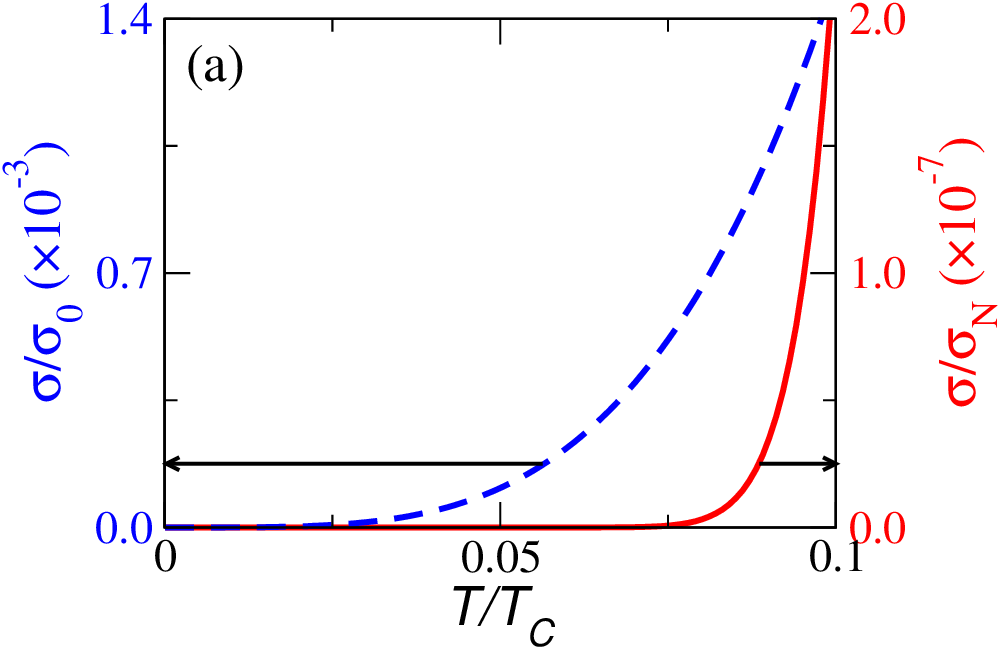}}\hspace{5mm}
\subfloat{\includegraphics[width=2.25in,clip]{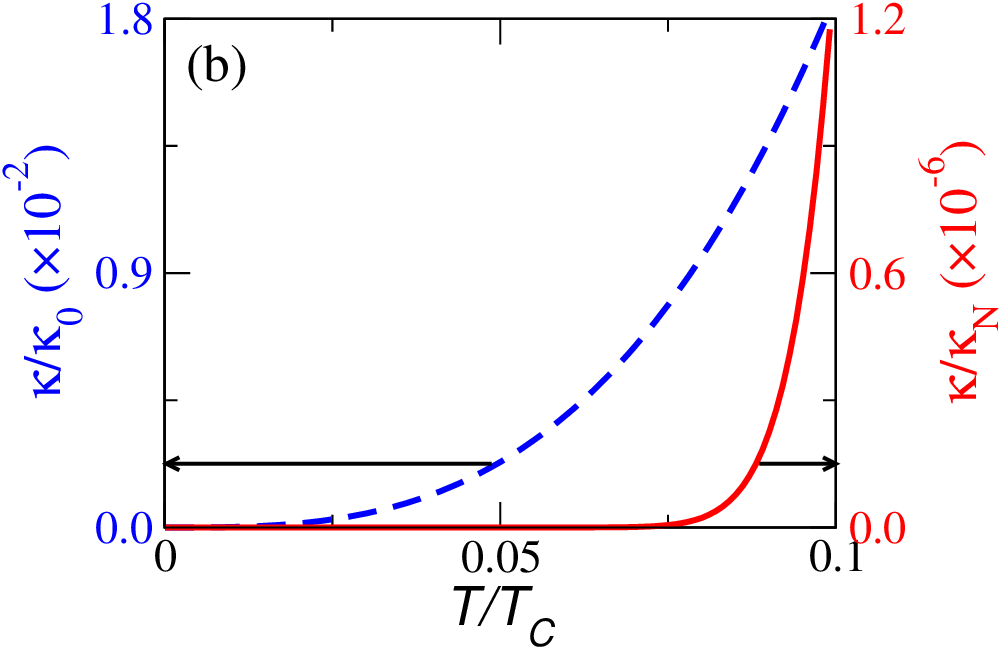}}
\caption{(Color online) The normalized (a) mass and (b) thermal  
conductivity coefficients for
bosonic (dashed blue line) and fermionic (solid red line) superfluids as functions of
$T/T_{C}$. The left and right axes are associated with the bosonic
and fermionic coefficients, respectively. The fermionic transport
coefficients are normalized by the normal state expressions 
$\sigma_{N}=\sigma(\Delta=0), \kappa_{N}=\kappa(\Delta=0)$.
In the bosonic case we define $T/T_{C}$=$T/\mu^{B}$. 
In the Bogoliubov approximation $n_{0}$ is the particle number at $T=0$. 
Thus our calculations are confined  to $T/T_{C}<<1$ and 
so we use the normalization $\sigma_{0}=\tfrac{(2m\mu^3)^{1/2}}{6\pi^2\gamma},
\kappa_{0}=\tfrac{(2m\mu^5)^{1/2}}{6\pi^2\gamma}$. Here the
mass conductivity corresponds to the $\omega \neq 0$ contribution.}
\label{fig:Transport}
\end{figure*}

\subsection{Bosonic One Loop Approximation and Correlation Functions}
In order to evaluate the various $\tensor{\chi}_{ij}$
we introduce the appropriate Green's functions.
These functions are well established for the case of fermionic BCS superfluids. 
For the bosonic case, the
one-loop approximation is based on the
Bogoliubov Green's functions and thus involves the Bogoliubov quasi-particle dispersion relation.
The Green's functions in the Bogoliubov approximation are 
\begin{align}
G(K)&=\frac{u_{\mb{k}}^2}{i\omega_{n}-E_{\mb{k}}}-\frac{v_{\mb{k}}^2}{i\omega_{n}+E_{\mb{k}}},\\
F(K)&=-u_{\mb{k}}v_{\mb{k}}\left(\frac{1}{i\omega_{n}-E_{\mb{k}}}-\frac{1}{i\omega_{n}+E_{\mb{k}}}\right),
\end{align}
where $u_{\mb{k}}^2=\tfrac{1}{2}(1+\xi_{\mb{k}}/E_{\mb{k}}),v_{\mb{k}}^2=u_{\mb{k}}^2-1.$
Because bosonic superfluid theories involve a controlled perturbation in the
interaction strength, they lead to a clear hierarchy of diagrams and
we can restrict attention in the dilute fluid limit to those involving one or
at most two Green's functions. The latter constitute the ``loops" of the
transport approximation.

For transverse response functions, 
the only diagrams that contribute are those that cannot be
divided into two parts by removing one line representing a single-particle propagator.
Such diagrams are called proper. \\\noindent The condensate contributions to a generic
correlation function (dependent on a single-particle Green's function)
are not proper, and therefore do not contribute to the transverse response functions.
It follows that, for a one-loop theory, the transverse component of a
generic correlation function is completely determined by diagrams containing only two single-particle Green's functions.
For longitudinal correlation functions, other than $L_{11}$, there are no condensate contributions and
again the leading order contribution involves two single-particle Green's functions.
In the superfluid phase there are two such Green's functions (the anomalous and normal Green's functions.)

At this bosonic one-loop level we
relate these correlation functions to
the imaginary time single particle Green's functions in position
space, given by
$G(x,\tau)$ (normal), $F(x,\tau)$ (anomalous), defined by:
$\langle T_{\tau}\psi(x_{1})
\psi^{+}(x_{2})\rangle=-G(x_{1}-x_{2})+n_{0}$ and $\langle
T_{\tau}\psi(x_{1})\psi(x_{2})\rangle=-F(x_{1}-x_{2})+n_{0}$.
For convenience, we make the following definitions:
the four vector summation $\sum_{K}\equiv-\tfrac{1}{\beta}\sum_{i\om_{n}}
\int\frac{d^{3}k}{(2\pi)^3}$, the vertex factors
$\mb{v}_{1}=\left(\tfrac{\mb{k}+\tfrac{1}{2}\mb{q}}{m}\right),
\mb{v}_{2}=\frac{\mb{q}}{2m},$ and
$\mb{v}_{3}=\left((i\om_{n}+i\om_{m})\frac{\mb{k}}{2m}+
i\om_{n}\frac{\mb{k+q}}{2m}\right)$. The dissipative parameter $\gamma$ previously introduced also serves to analytically
continue the Matsubara frequencies $i\om_{m}$ to
real frequencies $\om$ via: $i\om_{m}=\om+i\gamma$.

With these definitions, the
four momentum space correlation functions can be computed.
The particle current-particle current correlation function is
given by:
\begin{align}\label{eq:X11}
&\tensor{\chi}_{11}(\mb{q},i\om_{m})=n_{0}\mb{v}_{2}\mb{v}_{2}
[G(Q)+G(-Q)-F(Q)-F(-Q)]\nonumber\\&+\sum_{K}\mb{v}_{1}\mb{v}_{1}\left[G(K)G(K+Q)-F(K)F(K+Q)\right].
\end{align}
The particle current-heat current correlation function is:
\begin{align}\label{eq:X12}&\tensor{\chi}_{12}(\mb{q},i\om_{m})
=\sum_{K}\mb{v}_{1}\mb{v}_{3}\nonumber\\&\times\left[G(K)G(K+Q)+F(K)F(K+Q)
\right].\end{align}
The heat current-particle current correlation function is:
\begin{align}&\tensor{\chi}_{21}(\mb{q},i\om_{m})=\sum_{K}\mb{v}_{3}\mb{v}_{1}\nonumber\\&\times\left[G(K)G(K+Q)-F(K)F(K+Q)\right].
\end{align}
The heat current-heat current correlation function is:
\begin{align}\label{eq:X22}&\tensor{\chi}_{22}(\mb{q},i\om_{m})=\sum_{K}\mb{v}_{3}\mb{v}_{3}\nonumber\\&\times\left[G(K)G(K+Q)+F(K)F(K+Q)
\right].\end{align}
Our expressions in Eqs. (\ref{eq:X11}$-$\ref{eq:X22}) contain 
all possible contributions to the irreducible transverse response functions \cite{GriffinBookII}.
Note that, the correlation functions
$\tensor{\chi}_{12}$ and
$\tensor{\chi}_{21}$ differ in the relative sign of
the contribution from the anomalous Green's functions. 
It follows that, in order to satisfy the Onsager
relation, the anomalous Green's functions must give no contribution to
the transport coefficients $L_{12}$ and $L_{21}$. This is confirmed
explicitly by direct calculation.

As can be seen, the particle current-particle 
current correlation function ($\tensor\chi_{11}$) which appears in Eq. 
(\ref{eq:X11}), unlike all the other $\tensor\chi_{ij}$, contains a term proportional to the 
condensate density $n_{0}$. This term is purely longitudinal and of no
interest here. In order to ensure
charge conservation (via the longitudinal or $f$-sum rule)
in the superfluid phase, the condensate requires a consistent treatment,
analogous to collective mode effects in fermionic superfluids.

Finally, from the definitions of the transport coefficients, 
combined with the correlation functions in 
Eqs. (\ref{eq:X11}$-$\ref{eq:X22}) and the Bogoliubov 
Green's functions, the resulting transport coefficients 
$\eta^{B}$ and $\mathrm{Re}L^{B}_{ij}$ are given by 
the expressions in Eqs. (\ref{eq:NB}$-$\ref{eq:LB}).

Figure (\ref{fig:Transport}) shows the comparison between the
normalized low temperature bosonic and
fermionic transport coefficients, corresponding to mass and
thermal conductivity. It is clear from the
figures that the quasi-particle transport coefficients
at these low $T$ differ by several orders of magnitude. This is
due to the differences in the quasi-particle excitation
spectrum. From an experimental perspective, it appears
rather prohibitive to measure very low temperature transport properties of
Fermi systems. By contrast it appears Bose systems lend themselves to
these low $T$ studies.

\subsection{Low Temperature Analysis}
In general, the bosonic transport
coefficients exhibit power law behavior, whereas the
fermionic transport coefficients exhibit an
exponentially suppressed response. Explicitly, in the low temperature limits
($T<<\mu^{B},T_{C}$) we find that for bosons
\begin{align}
\mathrm{Re}L^{B}_{ij}&\rightarrow\frac{2\pi^2}{45\gamma}m^{1/2}(\mu^{B})^{i+j-9/2}T^{5-j},
\end{align}whereas for fermions
\begin{align}
\mathrm{Re}L^{F}_{11}&\rightarrow\frac{2g(E_{F})p_{F}^2}{3m^2\gamma}\left(\frac{2\pi\Delta_{0}}{T}\right)^{1/2}e^{-\Delta_{0}/T},\\
\mathrm{Re}L^{F}_{22}&\rightarrow\frac{2g(E_{F})p_{F}^2}{3m^2\gamma}\left(\frac{2\pi\Delta_{0}^3}{T}\right)^{1/2}e^{-\Delta_{0}/T},
\end{align}
where $p_{F}$ is the Fermi-momentum, $\Delta_{0}=\Delta(T\rightarrow0)$, and $g(E_{F})$ is the density of states at $E_F$. 
In BCS theory, assuming the chemical potential is of order 
$\mu^{F}\sim E_{F}$ and with exact particle-hole symmetry, $\mathrm{Re}L^{F}_{12}\rightarrow0$.

\begin{figure*}
\subfloat{\includegraphics[width=2.25in,clip]{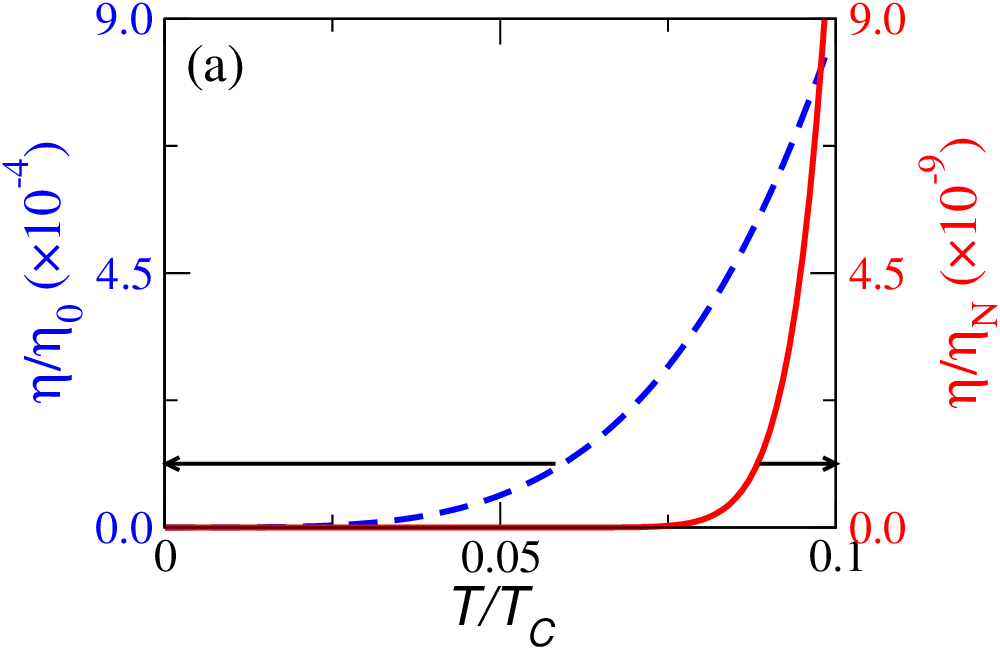}}\hspace{5mm}
\subfloat{\includegraphics[width=2.25in,clip]{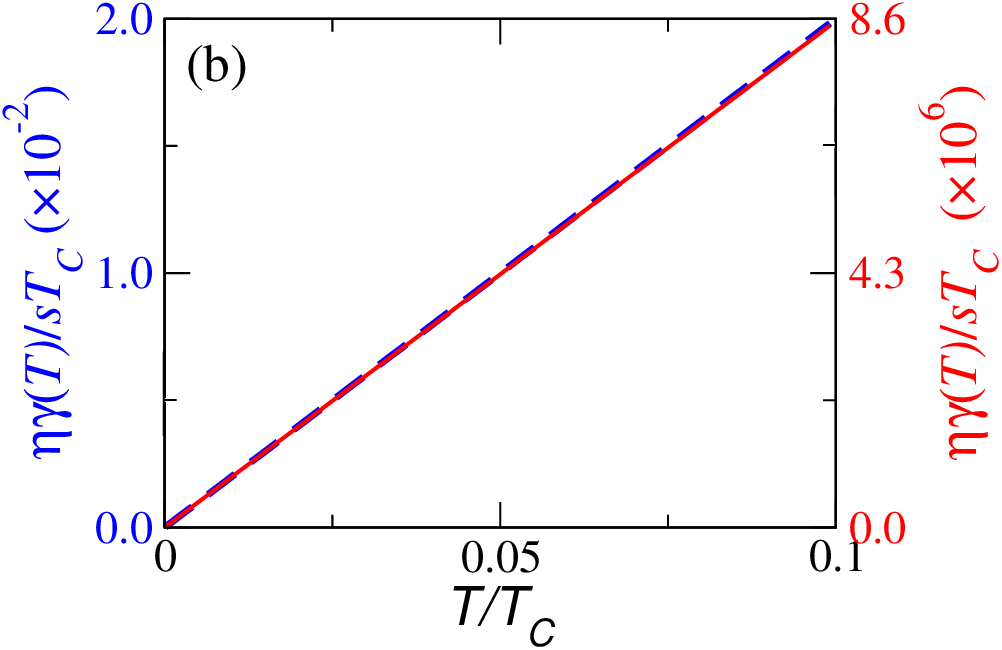}}
\caption{(Color online) (a) The  normalized shear viscosity of
bosonic (dashed blue line) and fermionic (solid red line) superfluids as functions of
$T/T_{C}$. The left and right axes are associated with the bosonic
and fermionic coefficients, respectively. The fermionic shear viscosity is normalized by the normal state expression $\eta_{N}=\eta(\Delta=0)$. In the bosonic case we define $T/T_{C}$=$T/\mu^{B}$. In the Bogoliubov approximation $n_{0}$ is the particle number at $T=0$. Thus our calculations are confined to $T/T_{C}<<1$ and so we use the normalization $\eta_{0}=\tfrac{(2m^3\mu^5)^{1/2}}{15\pi^2\gamma}$.
(b) The low temperature limit of the shear viscosity to entropy density ratio. Both limits are linear in $T/\gamma(T)$, and so have the same functional form. The only difference is the associated axes.}
\label{fig:Eta}
\end{figure*}

\section{Calculation of $\eta$ and $\eta/s$}
Kovtun, Son and Starinets (KSS) \cite{Kovtun} have made an interesting
conjecture concerning the shear viscosity. They conjecture that 
any relativistic quantum field theory at finite temperature and zero 
chemical potential has a ratio of shear viscosity to entropy density 
satisfying the bound $\eta/s\geq\hbar/(4\pi k_{B})$. Despite the 
construction of certain systems that violate the KSS bound 
\cite{Cherman}, the KSS conjecture has lead to renewed interest in
what the perfect fluids in nature are, i.e., those that come as 
close as possible to minimizing the conjectured bound. It has been 
shown by KSS that fluids that saturate this bound are those with a 
dual gravity description.

An interesting feature of the KSS bound is that it is independent of 
the speed of light $c$. Therefore, a non-relativistic quantum system 
is a possible candidate for the perfect fluid. Here we
investigate the magnitude of
$\eta/s$ arising from quasi-particle
transport in the bosonic one-loop and fermionic BCS superfluids.

A variant of the KSS conjecture extends the applicability of the conjectured bound of $\eta/s$ to
the case of non-zero chemical potential \cite{Son}. If we allow
$\mu^{B}, \mu^{F}\neq0$, then the low temperature entropy limits for bosons and fermions are
\begin{align}
\label{eq:SB}s^{B}&\rightarrow\frac{2\pi^2}{45}\left(\frac{m}{\mu^{B}}\right)^{3/2}T^3,\\
\label{eq:SF}s^{F}&\rightarrow 2g(E_{F})\left(\frac{2\pi\Delta_{0}^3}{T}\right)^{1/2}e^{-\Delta_{0}/T}.
\end{align}
Similarly the low temperature shear viscosity limits for bosons and fermions are
\begin{align}
\label{eq:EtaB}\eta^{B}&\rightarrow\frac{2\pi^2}{225\gamma}\left(\frac{m}{\mu^{B}}\right)^{3/2}T^4,\\
\label{eq:EtaF}\eta^{F}&\rightarrow\frac{2g(E_{F})p_{F}^4}{15m^2\gamma}\left(\frac{2\pi T}{\Delta_{0}}\right)^{1/2}e^{-\Delta_{0}/T}.
\end{align}
Depending on the temperature dependence of the quasi-particle lifetimes 
($\gamma^{-1}$) , the bosonic shear viscosity can exhibit 
an upturn for low temperatures. However, due to the exponentially 
suppressed term, the fermionic shear viscosity is not expected to exhibit an 
upturn, regardless of the parameter $\gamma(T)$.

Using the low temperature limits of $s$ and $\eta$ in Eqs. (\ref{eq:SB}$-$\ref{eq:EtaF}), 
we obtain the ratio $\eta/s$ for bosons and fermions:
\begin{align}\label{eq:ESB}\eta^{B}/s^{B}&\rightarrow\frac{1}{5}\frac{T}{\gamma},\\
\label{eq:ESF}\eta^{F}/s^{F}&\rightarrow\frac{4}{15}\left(\frac{E_{F}}{\Delta_{0}}\right)^2
\frac{T}{\gamma},\end{align} 
It should be noted that
once the entropy density is included, both bosons and fermions exhibit
the same $T/\gamma(T)$ dependence in their $\eta/s$ ratios.
This derivation confirms the arguments
given earlier, namely that systems with a quasi-particle description
have a large ratio of $\eta/s$, with the generic form $\eta/s\sim T/\gamma(T)$. 
An example of the temperature dependence used
for $\gamma$ in $^{4}$He \cite{Khalatnikov} would predict an upturn in
$\eta/s$ at low $T$ for the bosonic superfluid case.

While both bosonic and fermionic cases considered have a similar 
functional form for the ratio $\eta/s$, the low temperature limits 
of the entropy and shear viscosity of bosons and fermions are markedly different:

Figure (\ref{fig:Eta})
presents a plot of the normalized shear viscosity and the ratio $\eta/s$ for
bosonic and fermionic BCS superfluids.
While $\eta$ is highly suppressed for the fermionic case (as compared
with a bosonic superfluid), in the $\eta/s$ ratio the fermionic contribution
is highly enhanced. This is due to the fact that for fermions there are
two different energy scales present, $E_{F}$ and $\Delta_{0}$, while for bosons $\mu$ is the
only energy scale. Equation (\ref{eq:ESF}) shows that there is a factor of
the ratio of these two energy scales $(E_F/\Delta_0)^2$ which appears.
This is of course a very large number in the strict BCS limit, which
reflects the fact that the entropy in BCS theory is much smaller than
the shear viscosity.

\subsection{BCS-BEC Crossover}
\begin{figure*}
\includegraphics[width=5.5in,clip]{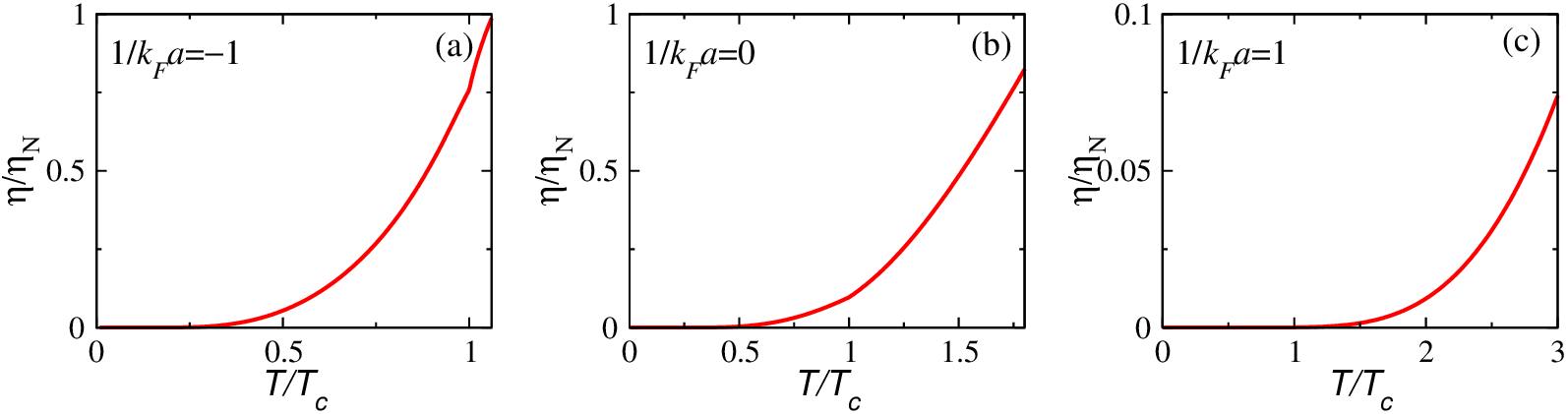}
\caption{(Color online) The normalized shear viscosity in BCS-BEC crossover theory. 
The shear viscosity is normalized by the normal state expression $\eta_{N}=\eta(T=T^{*})$. 
The parameters for each plot are (a) $T_{C}=0.12T_{F}, T^{*}=0.13T_{F}, 1/k_{F}a=-1$, 
(b) $T_{C}=0.26T_{F}, T^{*}=0.50T_{F}, 1/k_{F}a=0$, and (c) $T_{C}=0.21T_{F}, T^{*}=1.28T_{F}, 1/k_{F}a=1$.}
\label{fig:Eta_BCS_BEC}
\end{figure*}

Our primary analysis has been the transport properties of both bosonic and
fermionic superfluids in the temperature regime far below the critical temperature. 
While theories of bosonic superfluidity are restricted to low temperatures, 
fermionic superfluids can be studied up to $T_C$.
Equally interesting is the behavior in the normal phase ($T>T_{C}$)
of superfluids
in the presence of stronger attractive interactions, as associated with
the crossover from BCS to BEC.
In this normal or ``pseudogap" phase, pairing persists above $T_C$
and is expected to lead to suppressed shear viscosity.
With the discovery of the trapped atomic gases one has access to
fermionic superfluids with variable attraction, as parameterized
by the dimensionless scattering length $1/k_{F}a$, where
$1/k_{F}a=0$ is the so-called unitary regime.

One particular BCS-BEC scenario \cite{GuoII} by members of our
group (which is explicitly sum-rule consistent)
has addressed $\eta^{F}(T)$ theoretically over the entire
range of temperatures using as a framework the
BCS wave function for the ground state (with arbitrary attraction
and self consistent chemical potential). Here one finds
that $\eta^{F}=$
\begin{align}
&\int_{0}^{\infty}dk\ \frac{k^6}{15\pi^2 m^2}
\left(\frac{\xi_{\mb{k}}}{E_{\mb{k}}}\right)^2\left(1-\frac{\Delta^2_{pg}}{E_{\mb{k}}^2}\right)
\left(-\frac{\p f(E_{\mb{k}})}{\p E_{\mb{k}}}\right)\frac{1}{\gamma},
\end{align}
In this scenario there are two gap functions
$\Delta_{sc}$ and $\Delta_{pg}$, where the first represents the order
parameter, and the second the contribution to the excitation gap
associated with non-condensed pairs. The total gap is given by 
$\Delta^2=\Delta^{2}_{sc}+\Delta^{2}_{pg}$, so that the
excitation gap $E_{\mb{k}}$ takes the usual BCS form in terms of
the full gap $\Delta$. 
This previous work \cite{GuoII} and the above
equation shows that the effect of the non-condensed pairs, associated with 
the pseudogap $\Delta_{pg}$, is to reduce the shear viscosity. This is due to the fact that 
when pairs are present there are fewer fermions to contribute to the shear viscosity.

In Fig. (\ref{fig:Eta_BCS_BEC}) we show plots of the normalized shear viscosity
for this BCS-BEC crossover scenario. 
The plots from left to right correspond to passing from the BCS side
of unitarity, to unitarity, to the BEC side.
The exponential suppression of transport in fermionic superfluids is
then reflected in the behavior of $\eta$: as the gap $\Delta$ increases
in size from BCS to BEC the shear viscosity is accordingly
suppressed at the lowest $T$. It should be emphasized that the BEC limit
still reflects the pairing of fermions and will not coincide
with Bogoliubov descriptions of bosons, as the latter involves 
boson-boson interactions. These are not incorporated into the generalized
BCS wave function.

Of particular interest are the observations
\cite{ThomasI,ThomasII}
that the unitary superfluids have a shear viscosity which
is close to the KSS bound. This makes them very different
from the strict BCS superfluids with extremely large
$\eta/s$, which we studied earlier in this paper. There are hints from
Eq. (\ref{eq:ESF}) 
(which shows that $\eta/s$ for strict BCS fermions contains
a prefactor
$(T_{F}/T_{C})^2$) that as unitarity is approached and
$T_{F}/T_{C}$ becomes order one, that $\eta/s$ is significantly
reduced relative to the strict BCS case.
Indeed, the analogous prefactor $\sim(T_{F}/T_{C})^2$ is now several orders of magnitude smaller
than its counterpart for BCS theory.

As addressed in \cite{GuoII} and seen explicitly
in Fig. (\ref{fig:Eta_BCS_BEC}), the presence of the
non-condensed pairs via $\Delta_{pg}$ does not affect the
exponential suppression of $\eta$ at low temperatures. However,
at the same time the entropy acquires an additional bosonic
contribution 
($s=s^{F}+s^{B}$) \cite{Stajic}, where $s^{B}$ dominates at low $T$ and is a power of $T$.
Thus, the ratio $\eta/s$
will not be a linear function of $T/\gamma$, as was found for strict
BCS superfluids; rather, these near-unitary superfluids will exhibit
near perfect fluidity.

We summarize this discussion by emphasizing that our theory of
BCS-BEC crossover is a
theory of fermions and the
BEC limit does not include the direct effects of
inter-boson interactions (which give rise to the sound
mode excitations of Bogoliubov theory). These sound-mode
effects do, of course, appear in the collective modes as
the Nambu Goldstone bosons. Such collective modes must
be included in some transport coefficients and must
not be included in others. More precisely the Nambu Goldstone
modes in fermionic superfluids
of this type couple to the longitudinal response. They
do not couple to the transverse response, of which the
shear viscosity is one example. Our previous work
on $\eta$ for the BCS-BEC system used a theoretical approach
which analytically satisfied the transverse sum rule \cite{GuoII}.
A failure to satisfy the sum rules
is one of the best internal checks on whether or not,
and precisely where,
collective modes must be included in a given response function.

\section{Conclusions} 
We have compared
the $\omega \rightarrow 0$ mass conductivity, the shear viscosity, and the thermal conductivity
in bosonic and fermionic superfluids based on a Kubo formula approach within 
the one-loop Bogoliubov and closely related
BCS approximation. At this level
of approximation, our work
demonstrates the formal (albeit non-quantitative)
similarity between the transport
behavior of both superfluid types. 
The transverse response functions 
do not contain condensate contributions. Similarly for
the longitudinal thermoelectric coefficients (aside from the
$\omega \equiv 0$ mass conductivity) no condensate
contributions appear.
Thus, it is appropriate to characterize these coefficients
entirely in terms of their quasi-particle contributions,
as we have done here.

Of central interest here is the fact that even though the
shear viscosity for Bogoliubov and BCS superfluids
have dramatically
different temperature dependence,
their ratios in terms of the entropy density have
precisely the same linear $T/\gamma(T)$ dependence (where $\gamma(T)$ is the inverse
quasi-particle lifetime) with
very different prefactors. 
When considering the extension of BCS theory to BCS-BEC
crossover near unitarity, we find a very different temperature
dependence. Here because there are both bosonic and fermionic
degrees of freedom, there is no simple $T/\gamma(T)$ scaling.
Indeed, due to the suppression of the shear viscosity, 
it appears that unitary Fermi gases are a candidate
for nearly perfect fluids.

We stress that
pure bosonic or Bogoliubov theories of superfluidity have a structure
not exhibited in the BCS-BEC crossover; this difference
arises due to the soft dispersion relation
present in the long wavelength limit of Bogoliubov theory.
Similarly, inter-boson interactions are not
directly present at the level of a BCS-based theory of unitarity. Here the dominant
many body physics is an attraction between fermions, as distinct
from boson-boson interactions. Recent experiments \cite{ThomasIII} seem to confirm
this exponential suppression in the low temperature
shear viscosity as unitarity is approached.

We end by noting that essentially all reasonable
models for the temperature dependence of the transport
lifetime will give an upturn in $\eta/s$ at low $T$, but not,
for the case of fermions, in $\eta$ itself. This
appears consistent with the observed differences between
helium-3 and helium-4 superfluids \cite{GuoI}.

This work is supported by NSF-MRSEC Grant
0820054. We thank Adam Ran\c con for many helpful
conversations.

\clearpage
\end{document}